\documentclass[prc,twocolumn,showpacs,floatfix,aps,10pt]{revtex4-1}

\usepackage{amsmath}
\usepackage{graphicx}

\begin{document}

\title{Monopole transition strength function of ${}^{12}$C  in three-$\alpha$ model}
\author{Souichi Ishikawa}  \email[E-mail:]{ishikawa@hosei.ac.jp}
\affiliation{
Science Research Center, Hosei University, 
2-17-1 Fujimi, Chiyoda, Tokyo 102-8160, Japan
} 

\date{\today}

\begin{abstract}
The energy level structure of ${}^{12}$C nucleus at a few MeV above the three-$\alpha$ threshold is still unsatisfactory known. 
For instance, most microscopic calculations predicted that there exist one $0^+$-state in this energy region besides the well known Hoyle state, while some experimental and theoretical studies show the existing of two $0^+$-states.

In this paper, I will take a three-$\alpha$-boson (3$\alpha$) model for bound and continuum  states in ${}^{12}$C, and study a transition process from the ${}^{12}$C($0_1^+$) ground state  to 3$\alpha$ $0^+$ continuum states by the electric monopole ($E0$) operator. 
The strength distribution of the process will be calculated as a function of $3\alpha$ energy using the Faddeev three-body theory. 
The Hamiltonian for the $3\alpha$ system consists of  two- and three-$\alpha$ potentials, and some three-$\alpha$ potentials with different range parameters will be examined.

Results of the strength function show a double-peaked bump at low energy region, which can be considered as two $0^+$-states.
The peak at higher energy may originate from a 3$\alpha$ resonant state. 
However, it is unlikely that the peak at the lower energy is related to a resonant state, which suggests that it may be due to  so called ``ghost anomaly''. 
Distributions of decaying particles are also calculated. 
\end{abstract}


\maketitle

{\em Introduction.}
The $\alpha$-particle is, because of its stiffness, considered as a constitution of many of nuclear states.  
Low-lying continuum states in ${}^{12}$C are interesting subjects to study as a $3\alpha$ system. 
A well known example is the first $0^{+}$ excited state ${}^{12}$C$(0_2^+)$ (the Hoyle state) at the excitation energy $E_x = 7.65$ MeV, or $E=0.38$ MeV, where $E$ is measured from the 3$\alpha$ threshold energy (7.27 MeV).
This state plays an essential role in the synthesis of  ${}^{12}$C from 3$\alpha$ continuum states in stars (the triple-$\alpha$ process) \cite{Ho54}. 

In spite of a long history of studies, however, there are still some uncertainties in the ${}^{12}$C level structure at low energies:
in the compilation of experimental data in 1990 \cite{Aj90}, a $0^+$-state was tentatively listed to exist at $E=3.0$ MeV.
Recently, some experimental evidences for the existence of a $2^+$-state in this energy region have been reported \cite{Fr09,It11,Fr12,Zi13}.   
In addition, the monopole strength function for the transition ${}^{12}\mathrm{C}(0_1^+) \to 3\alpha$ $0^+$-state extracted from ${}^{12}\mathrm{C}(\alpha,\alpha^\prime)$ reaction at $E_\alpha=386$ MeV indicates the existing of two $0^+$-states \cite{It11}.
Theoretically, some microscopic calculations of ${}^{12}$C \cite{Ue79, Ka80, Ka07, Ch07} predicted the existing of one $0^+$-state in this energy region, while  a semi-microscopic model (the orthogonal-condition model) combined with the complex scaling method \cite{Ku05,Ku07,Oh13} as well as recent microscopic calculations \cite{Fu15, Zh16} result  two $0^{+}$-states to exist. 
 
Refs. \cite{Ku05,Ku07,Oh13} demonstrate that a 3$\alpha$ Hamiltonian possesses two complex eigenvalues of the form that $E_r - \frac{\imath}{2} \Gamma$, which can contribute to  the strength function  as a resonant state at $E=E_r$ of the width $\Gamma$.
Because of large values of the width for the both states, about 1.5 MeV, it is not clear what effects of these poles appear in the strength function at the real energy. 

In this paper, I will take a $3\alpha$ model for ${}^{12}$C and study a transition of the ${}^{12}$C ground state ${}^{12}\mathrm{C}(0_1^+)$ to 3$\alpha$ $0^+$-states in continuum by the $E0$ operator without a bound state approximation. 
This is an extension of recent works \cite{Is13,Is14}, in which $3\alpha$ continuum states up to $E=0.6$ MeV were studied to calculate the reaction rate of the triple-$\alpha$ process at stellar temperature $T\sim10^9$ K, and to clarify the decay mode of the Hoyle state. 
The $3\alpha$ system has been one of challenging few-body problems because of its mathematical difficulty in  treating the long-range Coulomb force among the $\alpha$-particles.
However, it is just remarked that recent calculations of the triple-$\alpha$ reaction rate \cite{Ak15,Su16} agree with that of the present method \cite{Is13} reasonably, which guarantees the accuracy of the calculations in a practical level.

In the following, the formalism to calculate the transition strength function (TSF) \cite{Is13,Is14} will be briefly described.  
Then after introducing some interaction models among $\alpha$-particles,  results for the TSF and  energy spectrum of decaying $\alpha$-particles will be presented.
Finally, summary will be given.

{\em Formalism.}
Let us consider a transition process of a 3$\alpha$ bound state $\vert \Psi_b \rangle$ to 3$\alpha$ continuum states of energy $E$ that is induced by the $E0$ operator $\hat{O}_{E0}$. Here, $\hat{O}_{E0}$ is written as follows (see, e.g. \cite{Ya08}),
\begin{equation}
\hat{O}_{E0}  = \boldsymbol{x}^2 + \frac{4}{3}\boldsymbol{y}^2,
\label{eq:12C_gs-E2-3a-2}
\end{equation}
where  $\boldsymbol{x}$ is  the relative coordinate of an $\alpha$-pair and $\boldsymbol{y}$ is the relative coordinate of the third (spectator) $\alpha$-particle with respect to the center of mass (c.m.) of the pair.

The final 3$\alpha$ state is characterized by the Jacobi momenta $\boldsymbol{q}$ and $\boldsymbol{p}$,
\begin{eqnarray}
\boldsymbol{q} &=&  \frac12 \left( \boldsymbol{k}_1 - \boldsymbol{k}_2 \right), 
\cr
\boldsymbol{p} &=& \frac23 \boldsymbol{k}_3 - \frac13 \left( \boldsymbol{k}_1 + \boldsymbol{k}_2 \right) = \boldsymbol{k}_3,
\label{eq:qp-k_i}
\end{eqnarray}
satisfying the energy conservation law,
\begin{equation}
\frac{1}{m_\alpha}\boldsymbol{q}^2 + \frac{3}{4m_\alpha}\boldsymbol{p}^2 = E_q + E_p= E, 
\label{eq:EqEp}
\end{equation}
where $\boldsymbol{k}_i$ is the momentum of the $i$-th $\alpha$-particle in the 3$\alpha$ c.m. system, and $m_\alpha$ is the $\alpha$ particle mass.

The $E0$-TSF is defined by 
\begin{eqnarray}
S_{E0}(E)  
&=&
\int d\boldsymbol{q} d\boldsymbol{p} 
\left\vert \langle \Psi^{(-)}_{\boldsymbol{q}\boldsymbol{p}} (E)\vert \hat{O}_{E0} \vert  \Psi_b \rangle  \right\vert^2
\delta\left(E- (E_q+E_p) \right)
\cr
&=&
\frac{m_\alpha^{2}}{3}  
\int d\hat{\boldsymbol{q}} d\hat{\boldsymbol{p}}  p q dE_q
\left\vert \langle \Psi^{(-)}_{\boldsymbol{q}\boldsymbol{p}} (E)\vert \hat{O}_{E0} \vert  \Psi_b \rangle  \right\vert^2,
\label{eq:SE_psi}
\end{eqnarray}
where $\vert \Psi^{(-)}_{\boldsymbol{q}\boldsymbol{p}}(E) \rangle$ is an eigenstate of the 3$\alpha$ Hamiltonian $H_{3\alpha}$ with incoming wave boundary condition, 
To calculate $S_{E0}(E)$, here, let us define a wave function for the transition process:
\begin{equation}
\left\vert \Psi (E) \right\rangle
= 
\frac{1}{E + \imath \epsilon - H_{3\alpha}} \hat{O}_{E0} \left\vert \Psi_b \right\rangle.
\label{eq:Psi_def}
\end{equation}

The wave function $\Psi (E)$  in Eq. (\ref{eq:Psi_def}) is obtained by applying the Faddeev three-body formalism \cite{Fa61} and a technique to solve integral equations in coordinate space with accommodating the long range Coulomb force effects. 
(See Refs. \cite{Is13,Is14,Is09} for the details of the calculations.)
This procedure provides the wave function $\Psi(E)$ as well as a three-body breakup amplitude $F^{(B)}(\hat{\boldsymbol{q}},\hat{\boldsymbol{p}},E_q;E)$ that is related to the transition amplitude as
\begin{eqnarray}
&&F^{(B)}(\hat{\boldsymbol{q}}, \hat{\boldsymbol{p}}, E_q;E) 
\cr
&=& 
  e^{\frac{\pi}{4}\imath} \sqrt{\frac{\pi}{2}}  \left(\frac{4}{3}\right)^{3/2}
  \frac{m_\alpha^{7/4} {E^{3/4}} }{\hbar^{1/2}} 
  \langle \Psi^{(-)}_{\boldsymbol{q}\boldsymbol{p}}(E) \vert \hat{O}_{E0} \vert  \Psi_b \rangle.
\label{eq:F_B-amp}
\end{eqnarray}

As Eq.  (\ref{eq:Psi_def}) indicates, the function $\Psi (E)$ has a purely outgoing wave in asymptotic region.
Distributions of outgoing particles are obtained by a current flux calculated from  the amplitude $F^{(B)}(\hat{\boldsymbol{q}}, \hat{\boldsymbol{p}}, E_q;E)$.
Explicitly, the flux of the particles for a configuration that  
$\hat{\boldsymbol{q}} \sim \hat{\boldsymbol{q}}+d\hat{\boldsymbol{q}}$, 
$\hat{\boldsymbol{p}} \sim \hat{\boldsymbol{p}}+d\hat{\boldsymbol{p}}$, and 
$E_{q} \sim E_{q}+ dE_{q}$, is given by 
\begin{eqnarray}
&& dJ(\hat{\boldsymbol{q}},\hat{\boldsymbol{p}},E_{q};E) 
\cr
&=&  
\left( \frac{3}{4}\right)^2 \frac{qp}{(m_\alpha E)^{3/2}}  
\left\vert F^{(B)}(\hat{\boldsymbol{q}},\hat{\boldsymbol{p}},E_{q};E) \right\vert^2
d\hat{\boldsymbol{q}} d\hat{\boldsymbol{p}} dE_{q}.
\label{eq:flux}
\end{eqnarray}

From  Eqs. (\ref{eq:SE_psi}), (\ref{eq:F_B-amp}), and (\ref{eq:flux}), one obtains,
\begin{equation}
S_{E0}(E) 
= \frac{\hbar}{2\pi} \int dJ(\hat{\boldsymbol{q}},\hat{\boldsymbol{p}},E_{q};E). 
\label{eq:SE_F}
\end{equation}
%

{\em Interaction models.}
For the $\alpha$-$\alpha$ interaction potential (2$\alpha$P),  I use  the model D of the Ali-Bodmer [AB(D)] potential \cite{Al66}  along  with the point $\alpha$-$\alpha$ Coulomb potential.
In solving $\Psi(E)$ in Eq. (\ref{eq:Psi_def}),  3$\alpha$ partial wave states with the angular momentum of the $\alpha$-$\alpha$ subsystem with 0, 2, and 4 are taken into account.

In addition, a 3$\alpha$ potential (3$\alpha$P) is included in the 3$\alpha$ Hamiltonian. 
In the present paper, I use the following functional form as in Refs. \cite{Fe96,Is13}, 
\begin{equation}
W_{3\alpha} =  W_{0} \exp\left[ -\frac{A_\alpha}{2b^2} \left(\boldsymbol{x}^2 + \frac{4}{3}\boldsymbol{y}^2\right) \right], 
\label{eq:VDel}
\end{equation}
where $A_\alpha=3.97$.

In view of uncertainties in current knowledge of  3$\alpha$P, some different values of the range parameter $b$ are examined, and then the strength parameter $W_0$ is determined to reproduce the Hoyle state energy. 
In Table \ref{tab:3aP-parameters-Delta}, chosen values of $b$ and thus determined $W_0$ are shown along with the calculated energy of the 3$\alpha$ bound state, $E[{}^{12}\text{C}(0_1^+)]$.
Since differences between the calculated energies and the experimental value of $E[{}^{12}\text{C}(0_1^+)]$ (-7.2746 MeV) \cite{Aj90} are not so large, the construction of a 3$\alpha$P that reproduces both of bound and continuum states in ${}^{12}$C quite well is left for a future problem. 
Hereafter, the 3$\alpha$P of the range parameter $b$ as well as calculations with it will be denoted by $\Delta(b)$.   
%

\begin{table}[t]
\caption{\label{tab:3aP-parameters-Delta}       
The range and strength parameters of the 3$\alpha$P models and calculated energy of 3$\alpha$ bound state corresponding to ${}^{12}\text{C}(0_1^+)$.
}
\begin{ruledtabular}
\begin{tabular}{ccc}
 $b$ (fm) &   $W_0$ (MeV) & $E[{}^{12}\text{C}(0_1^+)]$ (MeV) \\    
\hline
2.8  & -431.1 & -7.546  \\
3.0  &-303.66  & -7.584 \\ 
3.45   & -156.9  & -7.759  \\   
3.9 & -92.85  & -7.789 \\ 
4.6   & -48.54 & -7.281 \\ 
\end{tabular}
\end{ruledtabular}
\end{table}

\begin{figure}[th]
\begin{center}
\includegraphics[width=0.65\columnwidth,angle=-90]{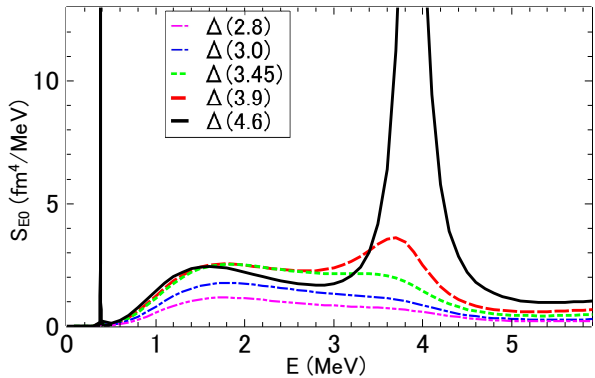}
\caption{\label{fig:be0-ab-d}
(Color online.) 
The $E0$-TSF of ${}^{12}$C as a function of the 3$\alpha$ energy $E$.
The peak value of $\Delta(4.6)$ is 26 fm$^4$/MeV at $E=3.9$ MeV.
}
\end{center}
\end{figure}

{\em $E0$ transition strength function.}
Results for $S_{E0}(E)$ are shown in Fig. \ref{fig:be0-ab-d}.  
One observes, besides a sharp peak of the Hoyle state at $E=0.38$ MeV, a continuous bump for $1~\mathrm{MeV} < E < 5$ MeV.
Two peaks at $E \approx 1.5$ MeV and  $E \approx4$ MeV are prominent for $\Delta(4.6)$, which is the longest-range 3$\alpha$P. 
The double-peak structure becomes inconspicuous as the 3$\alpha$P range becomes shorter. 
For a simplicity, I will call the peaks of $S_{E0}(E)$ at lower and higher energies as first and second peaks, respectively, in the following.
 
The monopole TSF extracted from ${}^{12}\mathrm{C}(\alpha,\alpha^\prime)$ reaction at $E_\alpha=386$ MeV  \cite{It11} reveals a double-peaked bump as a function of $E$, which suggests the existing of two $0^+$-states   at $E=1.77(9)$ MeV with the width $\Gamma=1.45(18)$ and $E=3.29(6)$ MeV with $\Gamma=1.42(8)$ MeV.  
The ratio of the peak height of the latter to the former is about 1.1.
Among the calculations in Fig. \ref{fig:be0-ab-d}, the $\Delta(3.9)$ is consistent with this tendency.

\begin{figure}[th]
\begin{center}
\includegraphics[width=0.65\columnwidth,angle=-90]{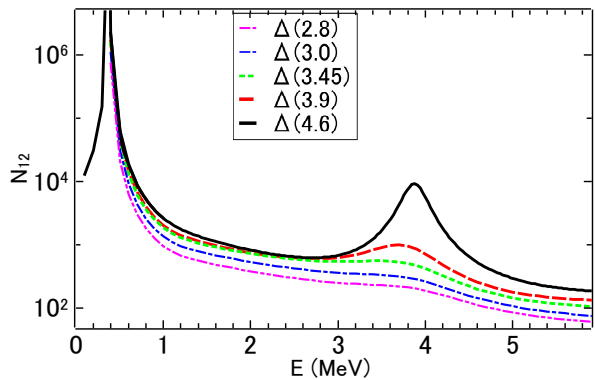}
\caption{\label{fig:fnorm-12}
(Color online.) 
The volume integral $N_{12}(E)$ defined in the text.
}
\end{center}
\end{figure}

Since the present calculations are performed at real energy $E$, it may be inadequate to judge definitively whether the peak of the TSF with a large width arises due to the resonant state or not.
As a possible information, I calculated a volume integral of the absolute square of the wave function $\Psi(E)$.
Since $\Psi(E)$ itself is not square normalizable, the integration range is restricted as $ x \le 12$ fm and $y \le 12$ fm resulting a finte value designated as $N_{12}(E)$.
The results are shown in Fig. \ref{fig:fnorm-12}.
The concentration of the amplitude at interior region matches the existence of a resonance, which is clear for the Hoyle state.
Another apparent peak in $N_{12}(E)$ is observed around $E=4$ MeV corresponding to the second peak.
However, no peak in $N_{12}(E)$ is observed around $E=1.5$ MeV, which indicates the first peak may not be caused by a resonant state.
One of possible explanations may be a ghost anomaly \cite{Ba62} associated with the Hoyle state.

No matter what the origin is, the calculated $E0$-TSF has a bump structure at $E \approx 3$ MeV.
Here, for a quantitative comparison, I introduce the $E0$ strength  for a certain range of the energy by  
\begin{equation}
B({E0};E_i,E_f) = \int_{E_i}^{E_f} S_{E0}(E) dE.
\label{eq:be0-strength}
\end{equation}
For the Hoyle state, which has a very narrow width [8.3(1.0) eV], the values: $E_i=0.37$ MeV and $E_f=0.39$ MeV, are used in Eq. (\ref{eq:be0-strength}), and the result will be denoted as $B({E0};\text{Hoyle})$. 
Since the two peaks in $S_{E0}(E)$ are not clearly separated for the  most of the calculations in Fig. \ref{fig:be0-ab-d}, I arbitrarily introduce two regions:  $(E_i, E_f)=(0.8,2.5)$ MeV and $(E_i, E_f)=(2.5,6.0)$ MeV.
Calculated values of the $E0$ strength in each region as well as the sum of them are shown in Table \ref{tab:calculated-BE0}.

\begin{table}[t]
\caption{\label{tab:calculated-BE0}       
The $E0$ strength, $B({E0};E_i,E_f)~ [\mathrm{fm}^4]$ defined in Eq. (\ref{eq:be0-strength}), calculated for some sets of $(E_i,E_f)$ as described in the text.
The energies are given in MeV.
}
\begin{ruledtabular}
\begin{tabular}{lcccc}
 & $(E_i,E_f)$ \\
Model   & (0.37,0.39) &  (0.8,2.5)  &  (2.5,6.0) & (0.8,6.0) \\
\hline
$\Delta(2.8)$  &  11.8  & 1.6 &1.8 &3.4 \\
$\Delta(3.0)$  & 17.6 & 2.4 & 2.6 & 5.0 \\
$\Delta(3.45)$  & 24.8   &3.5 & 4.3& 7.8 \\
$\Delta(3.9)$  & 28.9 & 3.6 & 6.0 & 9.6 \\
$\Delta(4.6)$  &  34.6  &3.3 & {14.2}&  {17.5} \\ 
\end{tabular}
\end{ruledtabular}
\end{table}

%
The microscopic calculations of the Hoyle state  \cite{Ue79, Ka80, Ka07, Ch07, Fu15, Zh16}  result  $40$ fm$^4$ to $45$ fm$^4$ for $B({E0};\text{Hoyle})$, which are larger than an experimental value deduced from inelastic electron scattering data: $29.9\pm1.0$ fm$^4$ \cite{Ch10}.
The present calculations of the $B(E0;\text{Hoyle})$ are smaller than those of the microscopic calculations, and one of them, namely $\Delta(3.9)$, is consistent with the experimental value.
 
In Ref. \cite{Cu13}, inelastic $(\alpha,\alpha^\prime)$ data are analyzed by a coupled-channel method, which gives $B(E0;\text{Hoyle})=20\pm4$ fm$^4$. 
This value is smaller than the experimental value of Ref.  \cite{Ch10}, but not so far from it. 
Also in Ref. \cite{Cu13}, the $E0$ strength with a Gaussian distribution centered at $E=3.3$ MeV and the width 3.0 MeV is optimized to fit the data, which results the strength $B(E0)=8.4\pm1.7~\text{fm}^4$.
If one consider this value may be comparable to $B(E0;0.8,6.0)$ in the present calculation, $\Delta(3.45)$ and $\Delta(3.9)$ are preferred.

It is noted that the 3$\alpha$P, Eq. (\ref{eq:VDel}), can be rewritten using the distance between particles $i$ and $j$, $r_{ij}$, as 
\begin{equation}
W_{3\alpha} =  W_{0} \exp\left(-\frac{r_{12}^2  + r_{23}^ 2+  r_{31}^2 }{a^2} \right)
\label{eq:VDel_2}
\end{equation}
with $a=\sqrt{\frac{3}{A_\alpha}}b$.
It is interesting that the values of $b=3.45$ fm and $b=3.9$ fm give $a=3.4$ fm and $a=3.0$ fm, respectively, and these correspond to a situation that two $\alpha$-particles are almost in touch.

{\em Relative energy distributions.}
In the transition process studied in this paper, three $\alpha$-particles finally spread out. 
The most important $3\alpha$ configuration in the final state should arise from  so called sequential decay (SD) process, in which three $\alpha$-particles are firstly separated to the 2$\alpha$ resonant state  [${}^{8}$Be$(0_1^+)$, $E_{r, 2\alpha}=92$ keV, $\Gamma_{2\alpha}=5.57(25)$ eV] and the rest $\alpha$-particle, and then the resonant state decays to two $\alpha$-particles. 
This configuration is identified by looking at the relative energy of two of particles, $i$ and $j$,
\begin{equation}
E_{ij} = \frac{1}{m_\alpha} \left(\frac{{\boldsymbol{k}_i-\boldsymbol{k}_j}}{2}\right)^2.
\label{eq:Eij}
\end{equation}
It is noted that these energies are not independent because of a kinematical condition, $E_{12}+E_{23}+E_{31}=\frac{3}{2}E$, which is obtained from Eqs. (\ref{eq:qp-k_i}), (\ref{eq:EqEp}), and (\ref{eq:Eij}).

In some experimental works measuring final three $\alpha$-particles decaying from $3\alpha$ continuum states \cite{Fr94,Ma12,Ki12,Ra13,It14}, the outgoing $\alpha$-particles are ordered by their energies in the c.m. system ($E_i=\frac{\boldsymbol{k}_i^2}{2m_\alpha}$) as $E_3 \ge E_1 \ge E_2$ in order to take into account the particle equivalence. 
The number of the events for the three particles within the phase space $d\hat{\boldsymbol{q}}d\hat{\boldsymbol{p}}dE_q$ is evaluated by the flux $dJ(\hat{\boldsymbol{q}},\hat{\boldsymbol{p}},E_q)$ of Eq. (\ref{eq:flux}) with this condition, where  the momenta $\boldsymbol{k}_i$ are related with  $\boldsymbol{q}$ and $\boldsymbol{p}$ by Eq. (\ref{eq:qp-k_i}).

\begin{figure}[tbh]
\begin{center}
\includegraphics[width=0.9\columnwidth]{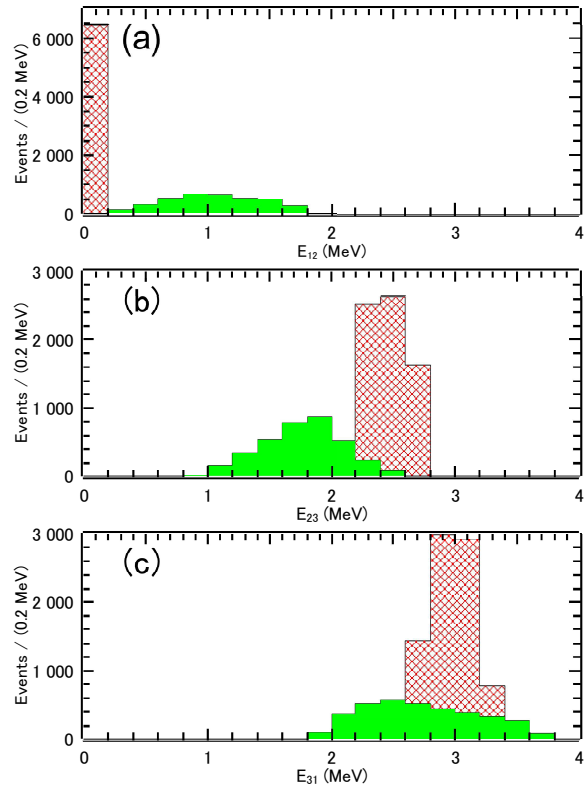}\hspace{2pc}%
\caption{\label{fig:rho-E12-E23-E31} 
(Color online.) 
The $\alpha$-pair relative energy distribution for $\Delta(3.9)$ at $E=3.7$ MeV.
The SD and non-SD contributions are expressed by hatched and filled areas, respectively.
}
\end{center}
\end{figure}

Distributions of the three $\alpha$-particles with respect to the relative energies of two $\alpha$-particles for $\Delta(3.9)$ at the energy of the second peak  ($E=3.7$ MeV) as an example, are shown in Fig. \ref{fig:rho-E12-E23-E31}. 
In the figure, the energy variables are divided to bins of the width = 0.2 MeV, and the numbers of the events included in the bins are plotted after normalizing the total number  to be $10^4$.
In Fig. \ref{fig:rho-E12-E23-E31} (a), the events of the lowest energy bin of $E_{12}$ (the hatched area)  correspond to the SD contribution. 
It is interesting that the non-SD events (the filled area) have a small peak at $E_{12} \approx 1$ MeV, which may caused by the ghost anomaly of $\alpha$-$\alpha$ state \cite{Ba62}.

The SD and non-SD events in Fig. \ref{fig:rho-E12-E23-E31} (a) are alternatively plotted as functions of $E_{23}$ and $E_{31}$ in Figs. \ref{fig:rho-E12-E23-E31} (b) and (c), respectively.
In the figures, the SD events are shown by hatched area and the non-SD events by filled area.

It is noted that the present 2$\alpha$P, AB(D), results a resonant state at $E=3.4$ MeV with the width 2.3 MeV for the angular momentum $L=2$ state, and  the peak energies of the distributions with respect to $E_{23}$ and $E_{31}$  are slightly less than this value.

\begin{figure}[tb]
\begin{center}
\includegraphics[width=0.65\columnwidth,angle=-90]{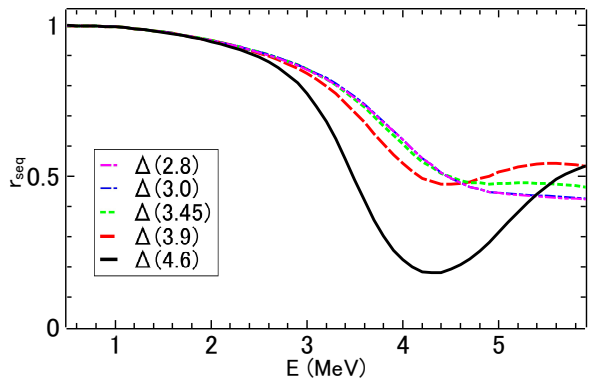}
\caption{\label{fig:be0-seq-ratio}
(Color online.) 
Ratio of the SD component to the total flux for the $E0$ transition of ${}^{12}\text{C}$ as a function of the $3\alpha$ energy $E$.
}
\end{center}
\end{figure}

Results for the ratio of numbers of the SD events to the total, $r_{\text{seq}}(E)$, are plotted in Fig. \ref{fig:be0-seq-ratio}. 
In Ref. \cite{Is13}, it is demonstrated that  $r_{\text{seq}}(E)$ exceeds 99 \% for the Hoyle state, which is consistent with experimental results of Refs. \cite{Fr94,Ma12,Ki12,Ra13,It14}.
This tendency holds for energy $E$ up to about 2 MeV.
For higher energy, the ratio reduced down to about 50 \%.

{\em Summary.}
Low energy 3$\alpha$ $0^+$ continuum states are studied by calculating the $E0$ transition strength function, $S_{E0}(E)$, of ${}^{12}$C in 3$\alpha$ model. 
The Ali-Bodmer model D $\alpha$-$\alpha$ potential together with 3$\alpha$Ps of some different ranges, whose parameters are determined to reproduce the resonance energy of the Hoyle state, are used for the $3\alpha$ calculations. 
For a 3$\alpha$P with longer range, calculated $S_{E0}(E)$ shows a bump at $E \approx 3$ MeV that has a double-peak structure.
As the range of 3$\alpha$P becomes shorter, the double-peak structure becomes less prominent.

From the behavior of the wave function that corresponds to the transition process, the second peak looks consistent with a resonant state.
On the other hand, the first peak does not show any resonance-like character.
This might be because of it occurs as the ghost anomaly associated with the Hoyle state.
The sequential decay process via ${}^{8}\mathrm{Be}(0_1^+)$ dominates for the lower-energy parts of the bump, but about 50 \% for the higher part.

Since these results rather strongly depend on the 3$\alpha$P, especially its range parameter, further experimental information of the $3\alpha$ transition as well as its $3\alpha$ decay mode is quite useful to reduce the uncertainties of interactions among the $\alpha$-particles.



\end{document}